\documentclass{aastex631}

\usepackage[whole]{bxcjkjatype}

\begin{document}

\title{ALMA 0.1~pc View of Molecular Clouds Associated with High-Mass Protostellar Systems in the Small Magellanic Cloud: Are Low-Metallicity Clouds Filamentary or Not?}

\author[0000-0002-2062-1600]{Kazuki Tokuda}
\affiliation{Department of Earth and Planetary Sciences, Faculty of Science, Kyushu University, Nishi-ku, Fukuoka 819-0395, Japan}
\affiliation{National Astronomical Observatory of Japan, National Institutes of Natural Sciences, 2-21-1 Osawa, Mitaka, Tokyo 181-8588, Japan}

\author[0000-0002-9627-1600]{Yuri Kunitoshi}
\affiliation{Department of Physics, Graduate School of Science, Osaka Metropolitan University, 1-1 Gakuen-cho, Naka-ku, Sakai, Osaka 599-8531, Japan}

\author[0000-0001-6149-1278]{Sarolta Zahorecz}
\affiliation{National Astronomical Observatory of Japan, National Institutes of Natural Sciences, 2-21-1 Osawa, Mitaka, Tokyo 181-8588, Japan}
\affiliation{Department of Physics, Graduate School of Science, Osaka Metropolitan University, 1-1 Gakuen-cho, Naka-ku, Sakai, Osaka 599-8531, Japan}

\author[0000-0002-6907-0926]{Kei E. I. Tanaka}
\affiliation{Department of Earth and Planetary Sciences, Institute of Science Tokyo, Meguro, Tokyo, 152-8551, Japan}

\author{Itsuki Murakoso}
\affiliation{Department of Earth and Planetary Sciences, Faculty of Science, Kyushu University, Nishi-ku, Fukuoka 819-0395, Japan}

\author[0000-0002-8217-7509]{Naoto Harada}
\affiliation{Department of Astronomy, Graduate School of Science, The University of Tokyo, 7-3-1 Hongo, Bunkyo-ku, Tokyo 113-0033, Japan}

\author[0000-0003-3990-1204]{Masato I. N. Kobayashi}
\affiliation{National Astronomical Observatory of Japan, National Institutes of Natural Sciences, 2-21-1 Osawa, Mitaka, Tokyo 181-8588, Japan}
\affiliation{I. Physikalisches Institut, Universit\"{a}t zu K\"{o}ln, Z\"{u}lpicher Str 77, D-50937 K\"{o}ln, Germany}

\author[0000-0002-7935-8771]{Tsuyoshi Inoue}
\affiliation{Department of Physics, Konan University, Okamoto 8-9-1, Kobe, Japan}

\author[0000-0003-2248-6032]{Marta Sewi{\l}o}
\affiliation{Exoplanets and Stellar Astrophysics Laboratory, NASA Goddard Space Flight Center, Greenbelt, MD 20771, USA}
\affiliation{Department of Astronomy, University of Maryland, College Park, MD 20742, USA}
\affiliation{Center for Research and Exploration in Space Science and Technology, NASA Goddard Space Flight Center, Greenbelt, MD 20771, USA}

\author[0000-0002-4098-8100]{Ayu Konishi}
\affiliation{Department of Physics, Graduate School of Science, Osaka Metropolitan University, 1-1 Gakuen-cho, Naka-ku, Sakai, Osaka 599-8531, Japan}

\author[0000-0002-0095-3624]{Takashi Shimonishi}
\affiliation{Institute of Science and Technology, Niigata University, Ikarashi-ninocho 8050, Nishi-ku, Niigata 950-2181, Japan}

\author[0000-0001-7511-0034]{Yichen Zhang}
\affiliation{Department of Astronomy, Shanghai Jiao Tong University, 800 Dongchuan Rd., Minhang, Shanghai 200240, China}

\author[0000-0002-8966-9856]{Yasuo Fukui}
\affiliation{Department of Physics, Nagoya University, Furo-cho, Chikusa-ku, Nagoya 464-8601, Japan}

\author[0000-0001-7813-0380]{Akiko Kawamura}
\affiliation{National Astronomical Observatory of Japan, National Institutes of Natural Sciences, 2-21-1 Osawa, Mitaka, Tokyo 181-8588, Japan}

\author[0000-0001-7826-3837]{Toshikazu Onishi}
\affiliation{Department of Physics, Graduate School of Science, Osaka Metropolitan University, 1-1 Gakuen-cho, Naka-ku, Sakai, Osaka 599-8531, Japan}

\author[0000-0002-0963-0872]{Masahiro N. Machida}
\affiliation{Department of Earth and Planetary Sciences, Faculty of Science, Kyushu University, Nishi-ku, Fukuoka 819-0395, Japan}

\begin{abstract}

Filamentary molecular clouds are an essential intermediate stage in the star formation process. To test whether these structures are universal throughout cosmic star formation history, it is crucial to study low-metallicity environments within the Local Group. We present an ALMA analysis of the ALMA archival data at the spatial resolution of $\sim$0.1\,pc for 17 massive young stellar objects (YSOs) in the Small Magellanic Cloud (SMC; Z $\sim$0.2\,$Z_{\odot}$). This sample represents approximately 30\% of the YSOs confirmed by Spitzer spectroscopy. Early ALMA studies of the SMC have shown that the CO emission line traces an H$_2$ number density of $\gtrsim$10$^4$\,cm$^{-3}$, an order of magnitude higher than in the typical Galactic environments. Using the CO($J$ = 3--2) data, we investigated the spatial and velocity distribution of molecular clouds. Our analysis shows that about 60\% of the clouds have steep radial profiles from the spine of the elongated structures, while the remaining clouds have a smooth distribution and are characterized by lower brightness temperatures. We categorized the former as filaments and the latter as non-filaments. Some of the filamentary clouds are associated with YSOs with outflows and exhibit higher temperatures, likely reflecting their formation conditions, suggesting that these clouds are younger than non-filamentary ones. This indicates that even if filaments form during star formation, their steep structures may become less prominent and transit to a lower-temperature state. Such transitions in structure and temperature have not been reported in metal-rich regions, highlighting a key behavior for characterizing the evolution of the interstellar medium and star formation in low-metallicity environments.

\end{abstract}


\section{Introduction} \label{sec:intro}

Understanding star formation in low-metallicity environments is crucial for gaining insights into the early stages of cosmic star formation. The Small Magellanic Cloud (SMC), a star-forming galaxy within the Local Group, presents an ideal target for such studies thanks to its significantly sub-solar metallicity (Z $\sim$0.2\,$Z_{\odot}$; \citealt{Russell_1992,Rolleston_1999,Pagel_2003}). The relatively low metallicity of the SMC mimics conditions of about 10 billion years ago, when the cosmic star formation rate was at its peak \citep{Pei_1999,Madau_2014}. Moreover, given its close proximity of roughly 62\,kpc \citep{Graczyk_2020}, the SMC offers a unique laboratory for investigating how star formation unfolds in environments similar to those prevalent in the early universe.

Generally, in low-metallicity environments, heating and cooling efficiencies differ from those in the Milky Way(MW)-like metal-rich regions \citep[e.g.,][]{Omukai_2005,Inoue_2015,Kobayashi_2023}, meaning that the processes that form and sustain molecular clouds and their internal structures are not guaranteed to be the same. In low-metallicity environments, it is not obvious whether filamentary molecular clouds, which are commonly reported across star-forming regions in the solar neighborhood \citep[e.g.,][]{Mizuno_1995,Nagahama_1998,Onishi_1999,Andre_2014}, would form and be maintained in the same manner \citep{Chon_2021}. If thermal evolution and the accompanying structural evolution depend on metallicity, they could also influence the behavior of the Initial Mass Function (IMF) throughout cosmic history. In low-metallicity environments, it is suggested that the high-temperature gas increases the Jeans mass, leading to the formation of low-mass stars that might be less efficient \cite[e.g.,][]{Omukai_2005}. On the other hand, observational studies on the IMF in the SMC or outskirts of the MW have reported conflicting results; some suggest no significant difference from the standard slope  \citep{Chiosi_2007}, while others indicate a top-heavy (or bottom-light) mass function \cite[e.g.,][]{Kalirai_2013,
Yasui_2023}. In either case, the relationship between the formed stars and their progenitor molecular clouds is not as well understood observationally as it is in the solar neighborhood.

Extensive observations of dense molecular material, the progenitors of star formation, have been conducted using various instruments ranging from small- to large-aperture millimeter and submillimeter single-dish telescopes \citep[e.g.,][]{Rubio_1991,Mizuno_2001,Bolatto_2003,Muller_2010,Saldano_2023,Saldano_2024}. 
Interferometric ALMA (Atacama Large Millimeter/submillimeter Array) observations \citep[e.g.,][]{Muraoka_2017,Jameson_2018,Neelamkodan_2021,Tokuda_2021,Oneill_2022,Ohno_2023,Tarantino_2024} and dust continuum imaging by space- and ground-based telescopes \citep[e.g.,][]{Gordon_2011,Gordon_2014,Jameson_2016,Takekoshi_2017,Takekoshi_2018} are also playing significant roles in revealing the nature of molecular clouds in the SMC. Although there are some common features with molecular clouds in the MW and the Large Magellanic Cloud (LMC) with a distance of 50\,pc \citep{Pietrzynski_2019}, the SMC exhibits unique characteristics. ALMA molecular gas observations, combined with other diffuse gas tracers, have revealed that the bright CO emission is visible in regions of high H$_2$ column density \citep{Jameson_2018}. The multiline analysis of CO and its isotopes estimated that the H$_2$ number density of the SMC clouds is $\sim$10$^4$\,cm$^{-3}$, and their kinematic temperatures are 30--50\,K \citep{Muraoka_2017}, which are several times to an order of magnitude higher than those derived from CO observations in the famous giant molecular clouds, Orion-A/B in the MW \citep{Nishimura_2015}. Recently, CO outflows have been detected from massive protostars in the SMC \citep{Tokuda_2022b,Shimonishi_2023}. Because protostellar outflows were discovered much earlier in the LMC \citep{Fukui_2015,Shimonishi_2016}, it was expected that the SMC would yield similar findings soon afterward. In reality, however, these discoveries were delayed by several years. One reason for this delay is that CO can only trace very high-density flow, leading to a lower detection probability than in the LMC \citep{Tokuda_2022b}. A large-scale survey with the Atacama Compact Array (ACA) at a spatial resolution of $\sim$2\,pc demonstrated that many CO clouds are comparable to the beam size. As a whole, more than 90\% of H$_2$ molecules are in the CO-dark phase, supporting the idea that even low-$J$ transitions of CO serve to provide deeper insight into molecular clouds. The molecular clouds have diverse morphologies, including filamentary \citep{Neelamkodan_2021} and non-filamentary structures \citep{Muraoka_2017}. The previous studies primarily reported observations with resolutions of $\gtrsim$0.3\,pc, which are insufficient to see the typical 0.1\,pc width of filaments reported in the MW \citep[e.g.,][]{Arzounamian_2011,Andre_2016} and the LMC \citep{Tokuda_2019,Tokuda_2023}. This limitation has made it difficult to distinguish whether the observed diversity is intrinsic or an artifact of the observational resolution. 

In summary, a key objective is to elucidate the characteristics of molecular clouds associated with (high-mass) star formation in the SMC and to determine whether the cloud morphology is filamentary or not across various galactic environments. 
This will help to determine how common the star formation process is through filamentary clouds, which are believed to be quasi-universal structures in more metal-rich environments.

In this paper, Section~\ref{sec:obs} provides a description of the target sources and an overview of the ALMA observations and data reduction. Section~\ref{sec:res} presents the results, including the characterization of molecular clouds and the identification of filaments. The discussion of these findings is given in Section~\ref{sec:dis}, and the summary is provided in Section~\ref{sec:summary}.

\section{Target Descriptions, ALMA Observations, and Data Reductions} \label{sec:obs}

As explained in Section~\ref{sec:intro}, our primary objective is to elucidate the diversity of molecular cloud structures in the SMC. It is essential to analyze whether filamentary molecular clouds are present with a resolution as high as possible with a uniform observation setting across many samples. The following three projects: (1) \#2019.1.00534.S (P.I.: S. Zahorecz), (2) \#2021.1.00518 (P.I.: S. Zahorecz), and (3) \#2019.1.01770.S (P.I.: K. Tanaka), used the Band 7 (275--373 GHz) receiver to observe CO($J$ = 3--2, 345.79599\,GHz) emissions toward spectroscopically confirmed massive young stellar objects (YSOs; \citealt{Oliveira_2013}). These projects adopted similar frequency settings, with CO(3--2) observed at a frequency resolution of 0.564\,MHz (0.490\,km\,s$^{-1}$) with a bandwidth of 937.5\,MHz and 3840 channels for Projects (1) and (2), and 0.976\,MHz (0.848\,km\,s$^{-1}$) with a bandwidth of 937.5\,MHz and 1920 channels for Project (3). We mainly use the CO(3--2) data, but the measurement sets include additional high-density tracers, such as HCO$^+$(4--3, 356.73424\,GHz) (see \citealt{Tokuda_2022b,Tokuda_2023,Shimonishi_2023}). The frequency settings for the continuum observations are as follows: Projects (1) and (2) cover a total of 1.875\,GHz centered at 355.3\,GHz, while Project (3) allocates 1.875\,GHz each to three basebands centered at 346.7, 357.2, and 358.9\,GHz.

The configurations of the 12\,m array used during these observations are listed in Table~\ref{tab:target}, achieving a resolution of approximately 0\farcs2--0\farcs4, which corresponds to $\sim$0.1\,pc at the distance of the SMC, 62\,kpc \citep{Graczyk_2020}. Detailed descriptions of the observations are provided in \cite{Tokuda_2022b,Tokuda_2023,Shimonishi_2023}. The field of view was $\sim$ 20$\arcsec$, equivalent to 6\,pc, and the maximum recoverable scale (MRS) was $\sim$5$\arcsec$ (= 1.5\,pc), according to the ALMA proposal guide. We used the data from the 12\,m array alone for this study, which could lead to concerns of missing flux. However, Total Power (TP) data combined with ACA data of the YSO-associated CO clouds in the SMC indicated that their median radius is $\sim$1.5\,pc \citep{Ohno_2023}, which is comparable to MRS in the present study, making the missing flux issue less significant. For data reduction and imaging, we used the Common Astronomy Software Application package \citep{CASA_2022}, v5.6.1\UTF{2013}8. We utilized the \texttt{tclean} task with the multiscale deconvolver, applying the Briggs weighting with a robust parameter of 0.5. We created velocity cubes of CO(3--2) with a velocity bin of 0.5\,km\,s$^{-1}$. Since the data have a sufficiently high signal-to-noise ratio, the auto-masking technique described by \cite{Kepley2020} was used to select the emission mask during the CLEAN process. The resultant beam sizes and sensitivities ($T_{\rm rms}$) are listed in Table~\ref{tab:target}.

\begin{table}[htbp]
\centering
\caption{Observation Targets and Technical Details}
\label{tab:target}
\begin{tabular}{ccccccllcc}
\hline \hline
Name & \multicolumn{2}{c}{Coordinates\,(J2000)$^{\rm 1}$} & $L$$^{\rm 1}$ & Group$^{\rm 1}$ & Observation ID & Array config. & Beam size & $T_{\rm rms}$  & outflow$^{\rm 2}$ \\
& R.A. (hms) & Dec (dms) & $(10^4\,L_\odot)$ & & & & $\theta_{\rm maj} \times \theta_{\rm min}$(P.A.)  & (K) &\\
\hline
\#01 & 00:43:12.86 & -72:59:58.3 & 1.6 & PE  & 2019.1.01770.S & C43-3, C-3, C-4 & 0\farcs46 $\times$ 0\farcs36 ($36^\circ$)  & 1.0   & $\cdots$ \\
\#02 & 00:44:51.87 & -72:57:34.2 & 1.9 & S   & 2019.1.01770.S & C43-3, C-3, C-4 & 0\farcs46 $\times$ 0\farcs36 ($36^\circ$)   & 0.96  & $\cdots$\\
\#03 & 00:44:56.30 & -73:10:11.8 & 6.1 & S   & 2019.1.00534.S & C43-4           & 0\farcs43 $\times$ 0\farcs32 ($-4^\circ$)   & 0.63  & \checkmark\\
\#10 & 00:49:01.64 & -73:11:09.6 & 3.3 & P   & 2019.1.01770.S & C43-3, C-3, C-4 & 0\farcs46 $\times$ 0\farcs35 ($38^\circ$)   & 0.94  & $\cdots$\\
\#13 & 00:50:43.26 & -72:46:56.2 & 2.2 & PE  & 2019.1.00534.S & C43-4           & 0\farcs42 $\times$ 0\farcs32 ($-2^\circ$)   & 0.66  & $\cdots$\\
\#15 & 00:52:38.84 & -73:26:23.9 & 2.1 & PE  & 2021.1.00518.S & C-4, C-5        & 0\farcs32 $\times$ 0\farcs25 ($-39^\circ$)  & 0.81  & $\cdots$\\
\#16 & 00:53:25.36 & -72:42:53.2 & 1.2 & P   & 2021.1.00518.S & C-4, C-5        & 0\farcs31 $\times$ 0\farcs25 ($-40^\circ$)  & 0.99  & $\cdots$\\
\#17 & 00:54:02.31 & -73:21:18.6 & 2.2 & S   & 2019.1.00534.S & C43-4           & 0\farcs43 $\times$ 0\farcs32 ($-6^\circ$) & 0.54  & $\cdots$\\
\#18 & 00:54:03.36 & -73:19:38.4 & 2.8 & S   & 2019.1.00534.S & C43-4           & 0\farcs43 $\times$ 0\farcs32 ($-6^\circ$)   & 0.58  & \checkmark\\
\#21 & 00:56:06.50 & -72:47:22.7 & 1.1 & P   & 2021.1.00518.S & C-4, C-5        & 0\farcs32 $\times$ 0\farcs25 ($-40^\circ$)  & 0.84  & $\cdots$\\
\#23 & 00:58:06.41 & -72:04:07.3 & 1.4 & P   & 2021.1.00518.S & C-4, C-5        & 0\farcs31 $\times$ 0\farcs25 ($-42^\circ$)  & 0.85  & $\cdots$\\
\#25 & 01:01:31.70 & -71:50:40.3 & 1.7 & P   & 2021.1.00518.S & C-4, C-5        & 0\farcs31 $\times$ 0\farcs25 ($-42^\circ$)  & 0.82  & $\cdots$\\
\#26 & 01:02:48.54 & -71:53:18.0 & 1.2 & PE  & 2021.1.00518.S & C-4, C-5        & 0\farcs31 $\times$ 0\farcs24 ($-42^\circ$)  & 0.87  & $\cdots$\\
\#28 & 01:05:07.26 & -71:59:42.7 & 14.0& S   & 2019.1.00534.S & C43-4           & 0\farcs42 $\times$ 0\farcs32 ($-8^\circ$)   & 0.57  & $\cdots$\\
\#29 & 01:05:30.71 & -71:55:21.3 & 1.0 & S   & 2019.1.01770.S & C43-3, C-3, C-4 & 0\farcs44 $\times$ 0\farcs36 ($27^\circ$)   & 0.96  & $\cdots$\\
\#33 & 01:05:30.22 & -72:49:53.9 & 2.6 & S   & 2019.1.00534.S & C43-4           & 0\farcs43 $\times$ 0\farcs32 ($-7^\circ$)   & 0.62  & \checkmark\\
\#34 & 01:05:49.29 & -71:59:48.8 & 2.3 & S   & 2019.1.01770.S & C43-3, C-3, C-4 & 0\farcs44 $\times$ 0\farcs36 ($30^\circ$)   & 0.95  & \checkmark\\
\hline
\end{tabular}\\
\flushleft{{\bf Table notes:} [1] Catalog information from \cite{Oliveira_2013}. [2] High-velocity CO wings at the YSO position (see \cite{Shimonishi_2023} for \#03, \cite{Tokuda_2022b} for \#18, Appendix~\ref{A:COout_moms} for \#33 and \#34).}
\end{table}

\begin{figure}[htbp]
    \centering
    \includegraphics[width=0.7\columnwidth]{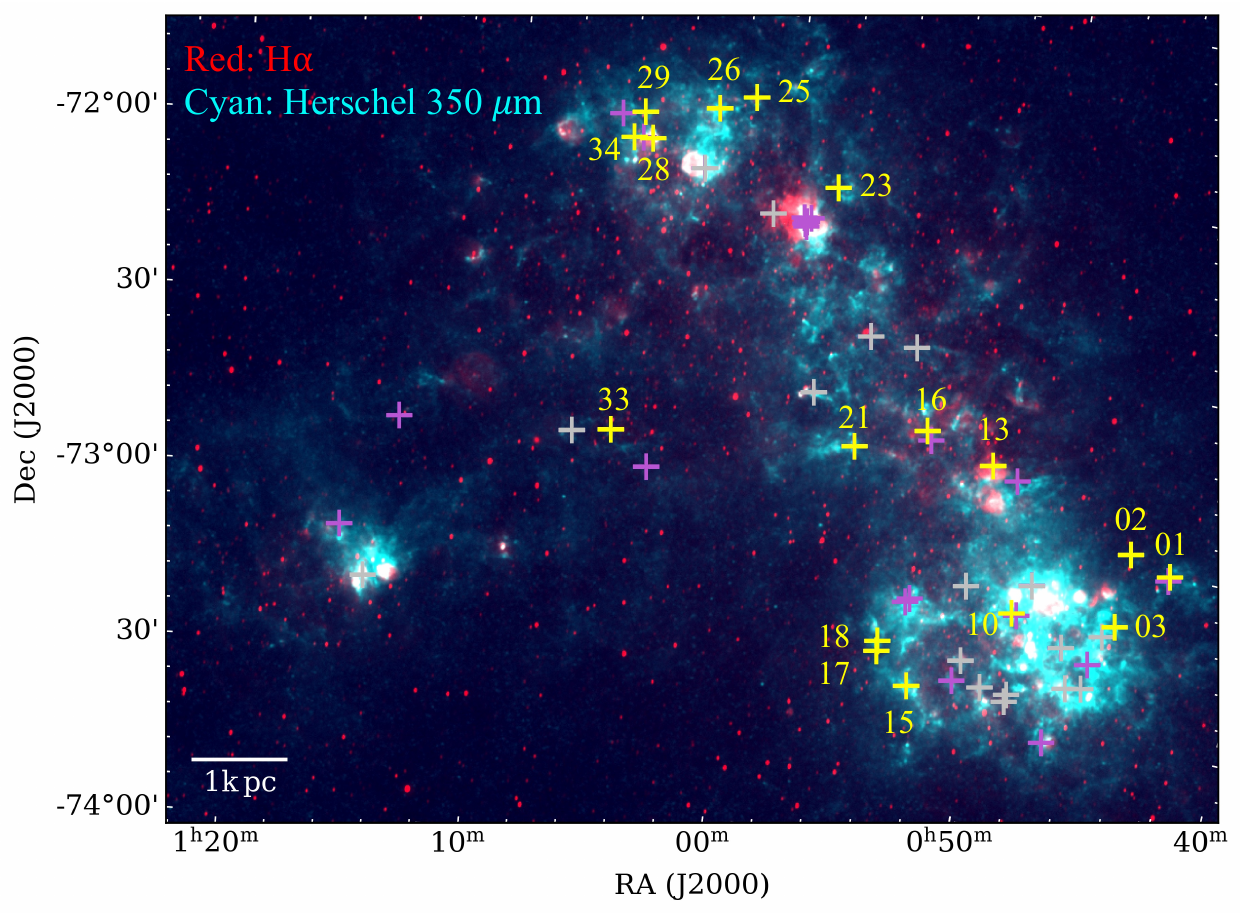}
    \caption{The overall view of the SMC and the positions of the target YSOs. The red and cyan backgrounds show H$\alpha$ \citep{Smith_1999} and Herschel 350\,$\mu$m \citep{Meixner_2013} images, respectively. The yellow and gray crosses indicate spectroscopically confirmed YSOs identified by \cite{Oliveira_2013}, with the former being the targets of this study, labeled with the corresponding numbers as shown in Table~\ref{tab:target}. The purple crosses represent additional spectroscopically confirmed YSOs identified by \cite{Ruffle_2015}.}
    \label{fig:SMC_Herscchel}
\end{figure}

After mining three data sets and removing duplicate targets, a total of 17 fields are available. In the SMC, \cite{Oliveira_2013} identified 33 YSOs with Spitzer spectroscopy, and a subsequent study by \cite{Ruffle_2015} confirmed an additional 19 sources, bringing the total to 51 YSOs. The original sample selection for this study is based on the former catalog alone, but our compiled ALMA sample covers approximately 30\% of the spectroscopically confirmed YSOs currently available in the SMC. The list of targeted YSOs and their properties, and the technical details of the observations are provided in Table~\ref{tab:target}. The positions of the YSOs from our sample are overlaid on the SMC image in Figure~\ref{fig:SMC_Herscchel}. The targets have luminosities exceeding 10$^{4}$\,$L_{\odot}$, indicating that they are likely to eventually evolve into high-mass stars. Table~\ref{tab:target} includes the $``$Group$"$ classification of the YSOs based on the Spitzer/IRS spectra introduced by \cite{Seale_2009}: Group~$``$S$"$ indicates deep absorption features predominantly due to dust and ices; $``$P$"$ denotes prominent PAH (polycyclic aromatic hydrocarbon) emission bands; and $``$PE$"$ represents a combination of strong PAH bands and ionized gas lines. These classifications were introduced to represent the evolutionary stages of YSOs by \cite{Seale_2009} for LMC spectroscopic YSOs, progressing from the youngest to the most evolved stages in the order of S, P, and PE. \cite{Oliveira_2013} classified the SMC sources according to the same criteria. Another category, $``$SE$"$, which lies between S and P, is not included in the targets of this study. 

To further characterize the YSOs based on the ALMA data, we identified CO outflows. First, we roughly determined the systemic velocity from the high-density gas tracer, HCO$^{+}$, at each protostar's position. Subsequently, we visually inspected the CO line profiles to search for the presence of wing emission extending up to $\pm$10\,km\,s$^{-1}$ in both redshifted and blueshifted sides, as detailed in \cite{Tokuda_2022b}. In addition to previously reported in YSO~\#03 and YSO~\#18 \citep{Tokuda_2022b,Shimonishi_2023}, we successfully detected new outflows in YSO~\#33 and YSO~\#34 (details are provided in the Appendix) as indicated in the last column in Table~\ref{tab:target}. The outflows identified in this study have characteristic sizes and maximum velocities on the order of 0.1\,pc and 10\,km\,s$^{-1}$, respectively, providing a dynamical time of approximately 10$^{4}$ years. These values are consistent with those reported in earlier studies toward YSO~\#03 and \#18 \citep[]{Tokuda_2022b, Shimonishi_2023}. While the dynamical time of these outflows does not necessarily represent the age of the protostars, the gas around the YSOs has not dissipated, suggesting that they are still in an early stage (see also Section~\ref{sec:res}). \cite{Ward_2017} performed near-infrared spectroscopic observations toward 10 YSOs from our samples using the Very Large Telescope (VLT). The sources with CO outflows detected by ALMA, except for YSO~\#18, tend to have higher accretion luminosities, as estimated from the Br$\gamma$ emission, compared to others.

\section{Results} \label{sec:res}

\subsection{Spatial distributions of $^{12}$CO(3--2) around massive YSOs in the SMC} \label{R:COimage}

\begin{figure}[htbp]
    \centering
    \includegraphics[width=1.0\columnwidth]{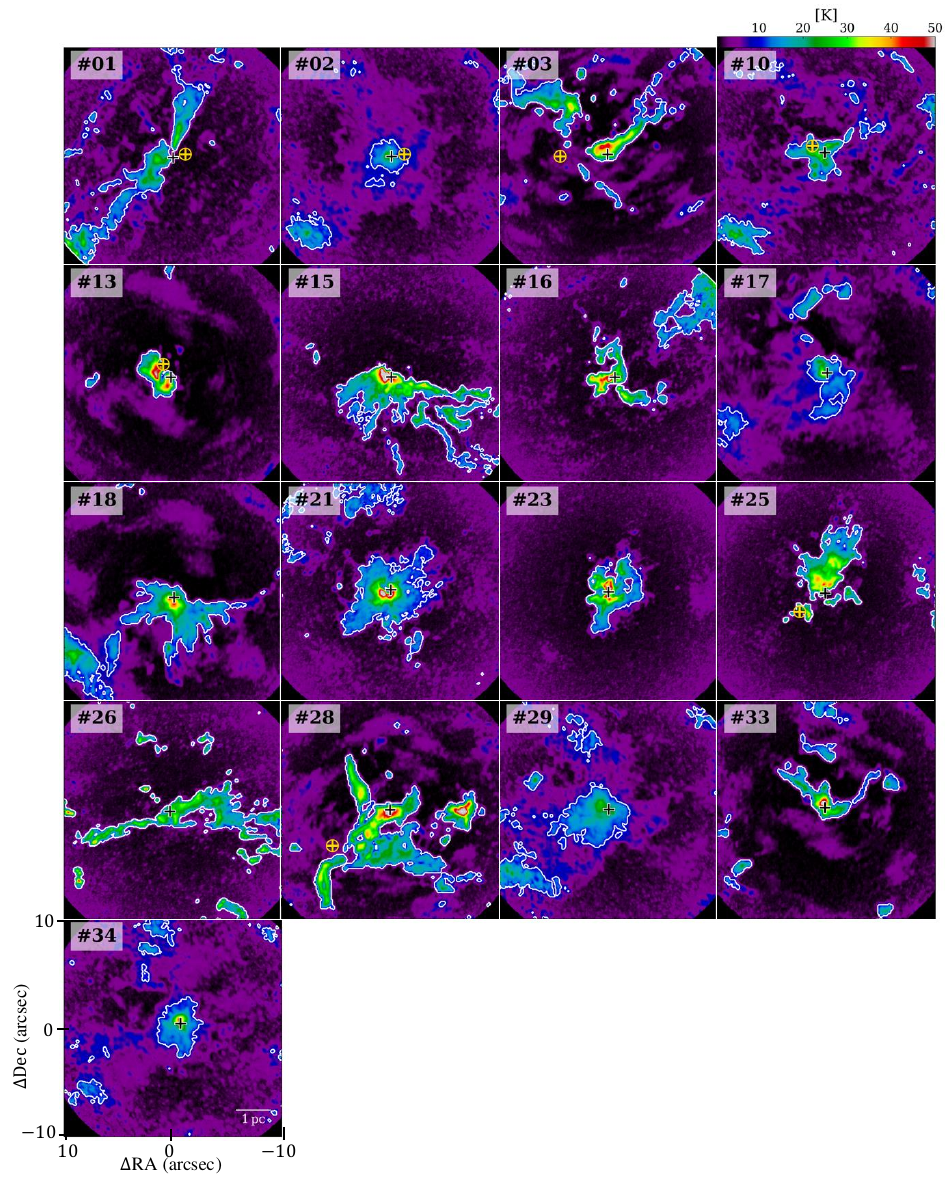}
    \caption{A gallery showing the spatial distribution of CO toward the observed ALMA fields. The color scale images represent the peak brightness temperature distribution of CO(3--2). Contours highlight regions where the CO intensity is 10\,K. The black crosses indicate the Spitzer positions of YSOs (see Table~\ref{tab:target}). The orange crosses surrounded by orange circles indicate the positions of compact H\;{\sc ii} regions traced by the radio-continuum emission \citep{Wong_2012}.}
    \label{fig:12CO32_Tpeak_view}
\end{figure}

Figure~\ref{fig:12CO32_Tpeak_view} shows the peak brightness temperature maps of CO(3--2) from the spectrum of each pixel for all observed ALMA fields. This study is the first attempt to observe cloud-scale structures in the SMC with a resolution close to 0.1\,pc, which is the highest currently reported in the literature, except for studies of dense cores associated with the infrared source IRAS~01042\UTF{2013}7215 \citep{Shimonishi_2018}
and protostellar outflow associated with Y246 \citep{Tokuda_2022b}, which corresponds to \#18 in this study. A variety of structures are present, ranging from filamentary to more diffuse configurations. Generally, the positions of the YSOs marked with black crosses in Figure~\ref{fig:12CO32_Tpeak_view} align well with the CO cloud peaks, indicating that the protostellar systems are in an early evolutionary phase before parental gas dissipation and/or consumption. Some fields also contain low-frequency radio continuum sources identified as point-like emission with a beam size of $\sim$5$\arcsec$ \citep{Wong_2012}, suggesting the presence of nearby ionizing sources or ionization from the YSO itself (\#01, \#02, \#13, \#25, and \#28). We derived the average peak brightness temperature ($T_{\rm peak}^{\rm ave}$) within the boundaries of regions enclosed by the contour level 10\,K that included the YSOs and listed them in Table~\ref{tab:filament}. For clouds that are close to the ionizing sources, the CO-derived temperatures range from 15 to 21\,K. Since the other sources show a similar temperature range, it suggests that the presence or absence of ionizing sources is not the primary factor determining the cloud temperature. A more detailed discussion of this topic is provided in Section~\ref{Dis:fil_form}.

\begin{table}[htbp]
\begin{center}
\caption{CO Cloud Properties}
\label{tab:filament}
\begin{tabular}{lccccccccccc}
\hline \hline
Name & $T_{\rm peak}^{\rm ave}$ & $l_{\rm sp,max}$ & $w_{\rm sp}$ &  $N_{\rm H_2}^{\rm ave}$ & $n_{\rm H_2}$ & $M_{\rm line}$ & $V_{\rm sys}$ & $\Delta v_{\rm FWHM}$ & $M^{\rm line}_{\rm vir}$ & $p$ & filament \\
 & (K) & (pc) & (pc)  & (cm$^{-2}$) & (cm$^{-3}$) & ($M_\odot$\,pc$^{-1}$) & (km\,s$^{-1}$) & (km\,s$^{-1}$) & ($M_\odot$\,pc$^{-1}$) & & \\
\hline
\#01 & 15 & 5.4 & 0.27 & 2.7e+23 & 3.2e+05 & 1.6e+03 & 111.2 & 1.6 & 2.0e+02 & $-$2.0  & \checkmark\\
\#02 & 14 & 1.5 & 0.19 & 2.9e+23 & 4.8e+05 & 1.2e+03 & 142.0 & 5.1 & 2.2e+03 & $-$0.6 &$\cdots$ \\
\#03 & 22 & 2.6 & 0.19 & 3.9e+23 & 6.6e+05 & 1.6e+03 & 132.5 & 3.7 & 1.2e+03 & $-$1.7  &\checkmark\\
\#10 & 17 & 4.0 & 0.17 & 7.0e+22 & 5.4e+05 & 9.9e+02 & 146.5 & 2.5 & 5.1e+02 & $-$1.0  & \checkmark\\
\#13 & 26 & $\cdots$   & $\cdots$&$\cdots$ &$\cdots$ &$\cdots$& 136.6 &$\cdots$ &$\cdots$ & $\cdots$  &   $\cdots$\\
\#15 & 21 & 6.1 & 0.20 & 4.5e+23 & 7.3e+05 & 1.3e+03 & 167.0 & 2.2 & 4.2e+02 & $-$1.0  & \checkmark\\
\#16 & 23 & 1.9 & 0.17 & 3.9e+23 & 7.4e+05 & 1.5e+03 & 144.0 & 1.8 & 2.8e+02 & $-$1.3  & \checkmark\\
\#17 & 14 & 7.5 & 0.34 & 2.0e+23 & 1.9e+05 & 1.6e+03 & 162.0 & 2.6 & 5.9e+02 & $-$0.4 & $\cdots$ \\
\#18 & 17 & 5.2 & 0.25 & 1.8e+23 & 2.3e+05 & 9.5e+02 & 163.6 & 2.7 & 6.3e+02 & $-$1.1  & \checkmark\\
\#21 & 16 & 4.2 & 0.25 & 4.1e+23 & 5.3e+05 & 2.3e+03 & 162.1 & 2.4 & 5.0e+02 & $-$0.7 & $\cdots$ \\
\#23 & 19 & 3.8 & 0.20 & 3.7e+23 & 6.0e+05 & 1.8e+03 & 156.8 & 2.0 & 3.4e+02 & $-$0.8  & $\cdots$\\
\#25 & 21 & 3.4 & 0.16 & 3.2e+23 & 6.5e+05 & 1.1e+03 & 159.3 & 2.0 & 3.3e+02 & $-$1.0  & \checkmark\\
\#26 & 19 & 7.6 & 0.16 & 3.2e+23 & 6.7e+05 & 1.1e+03 & 166.3 & 2.2 & 4.0e+02 & $-$1.2  & \checkmark\\
\#28 & 20 & 8.6 & 0.28 & 3.7e+23 & 4.3e+05 & 2.2e+03 & 189.3 & 2.5 & 5.4e+02 & $-$1.8  & \checkmark\\
\#29 & 15 & 6.8 & 0.18 & 2.7e+23 & 4.9e+05 & 1.1e+03 & 165.0 & 3.6 & 1.1e+03 & $-$0.9 & $\cdots$ \\
\#33 & 21 & 2.6 & 0.22 & 3.2e+23 & 4.7e+05 & 1.6e+03 & 153.8 & 2.3 & 4.4e+02 & $-$1.6  &\checkmark\\
\#34 & 15 & 1.4 & 0.17 & 2.8e+23 & 8.0e+05 & 7.7e+02 & 190.7 & 6.0 & 3.0e+03 & $-$0.8 &$\cdots$ \\
\hline
\end{tabular}\\
\end{center}
\end{table}

\subsection{Characterization of CO Clouds} \label{R:filament_id}

\begin{figure}[htb!]
    \centering
    \includegraphics[width=1.0\columnwidth]{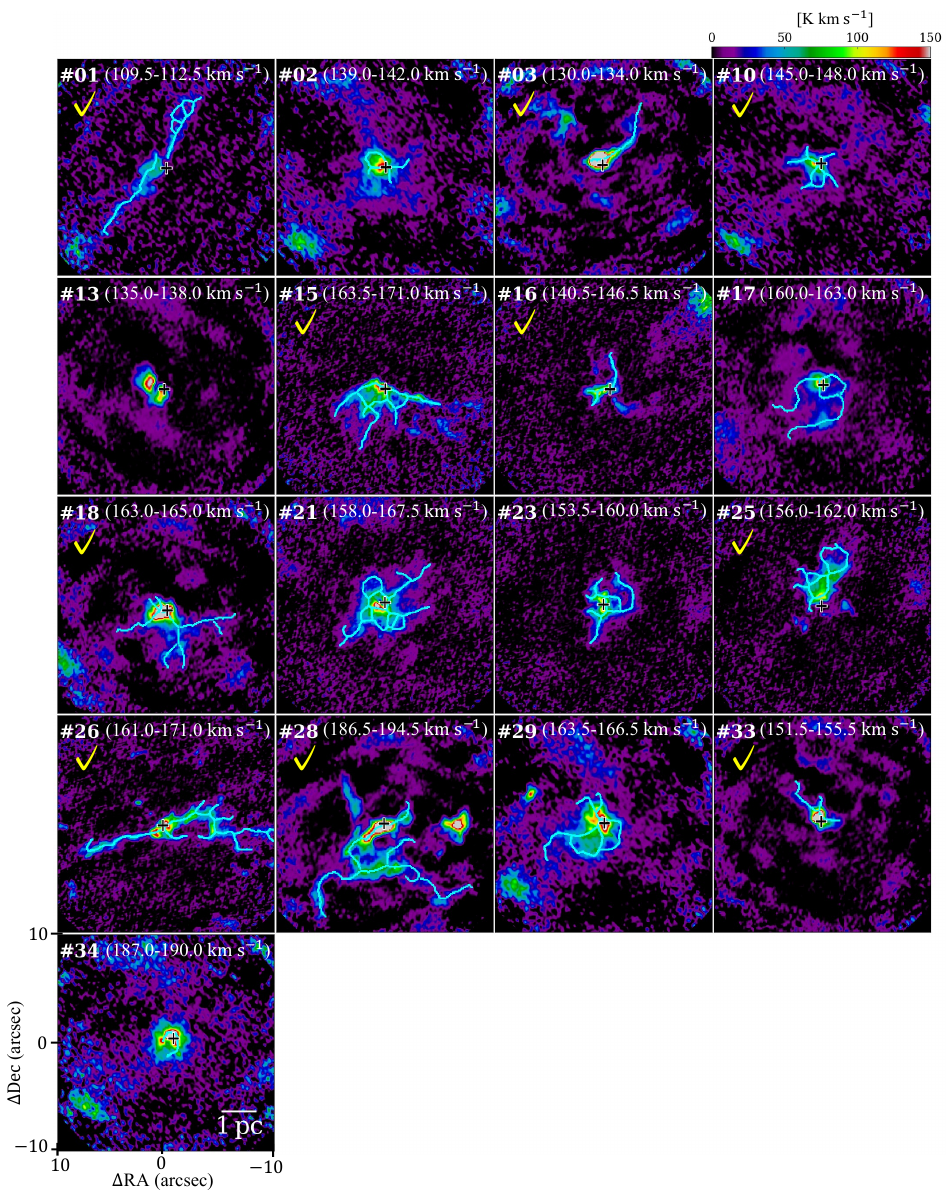}
    \caption{Moment~0 maps of CO(3--2) toward massive YSOs in the SMC. The integrated-velocity ranges are provided at the top of each panel. Black plus signs represent the Spitzer positions of the YSOs \citep{Oliveira_2013}. Cyan lines show spines identified in the FilFinder analysis. Those marked with a yellow check mark in the top left corner have steep radial profiles and are categorized as filaments (see Section~\ref{R:filament_id}).}
    \label{fig:mom0}
\end{figure}

To further quantitatively characterize the properties of CO clouds, we searched for elongated structures toward all targets. We applied the FilFinder algorithm to extract elongated structures using uniform criteria on the velocity-integrated intensity (moment~0) images shown in Figure~\ref{fig:mom0}. In the initial phase, we performed an arctangent transform to flatten the image of bright compact features by a certain percentile value. The \texttt{flatten\_percent} parameter was set to 95. Subsequently, emission masks were created by applying an intensity threshold of 10--20\,K\,km\,s$^{-1}$, which is roughly scaled to the data noise level for each field, and selecting regions with an area exceeding five times the beam size. The other input parameters used were \texttt{adapt\_thresh} = 0.1\,pc and \texttt{smooth\_size} = 0.05\,pc. The identified spines are drawn in a gallery of all fields (Figure~\ref{fig:mom0}). 

Based on the FilFinder analysis results, we derived physical quantities listed in Table~\ref{tab:filament} using the following procedure. 
Along each spine, we measured $l_{\rm sp,max}$ and performed Gaussian fitting to the averaged profile perpendicular to the spine, determining the full width at half maximum (FWHM), denoted as $w_{\rm sp}$. The derived $w_{\rm sp}$ ranges from 0.17 to 0.34\,pc, with fitting errors typically $\pm$0.01--0.05\,pc. The H$_2$ column densities ($N_{\rm H_2}$) were derived by converting the CO(3--2) integrated intensities in the unit of K\,km\,s$^{-1}$, to CO(1--0) equivalents assuming a CO(3--2)/CO(1--0) ratio of 1. We then multiplied these values by an $X_{\rm CO}$ factor of 7.5 $\times$10$^{20}$\,cm$^{-2}$\,(K\,km\,s$^{-1}$)$^{-1}$ \citep{Muraoka_2017} to obtain the $N_{\rm H_2}$ for each pixel. Note that \cite{Saldano_2024} reported that the global average of the CO(3--2)/CO(1--0) ratio in the SMC is $\sim$0.7 derived from single-dish observations. The ratio in dense regions close to massive protostars remains not fully constrained; however, if we were to adopt such a lower ratio, the resulting column densities would increase accordingly. In addition, the $X_{\rm CO}$ is also not necessarily a stringent constraint, recent surveys justified these assumptions \citep{Tokuda_2021,Ohno_2023}, providing accuracy within a factor of 2--3. By dividing the average column density of the pixels along the spine, $N_{\rm H_2}^{\rm ave}$ by $w_{\rm sp}$, we determined the number density, $n_{\rm H_2}$. The $n_{\rm H_2}$ value is on the other of 10$^{5}$\,cm$^{-3}$, consistent with those suggested in the early multi-line study in the SMC \citep{Muraoka_2017} at a $\sim$1\,pc resolution, or even higher density owing to the higher resolution of our data. The identified spines in \#03 and \#18, for example, are spatially connected to the protostellar cores with an average density of $\sim$10$^{6}$\,cm$^{-3}$ \citep{Tokuda_2022b,Shimonishi_2023}, indirectly justifying the high densities traced by CO in these regions. \cite{Jameson_2018} estimated H$_2$ column densities in some of the SMC clouds to be on the order of 10$^{21}$\,cm$^{-2}$ based on the [C$\;${\sc ii}] and CO measurements at a resolution of $\sim$3\,pc. This is two orders of magnitude lower than our estimate, probably also due to the resolution effect rather than method differences. Using some of these derived values, we estimated the line mass of the spines, $M_{\rm line}$ ($=\mu m_{\rm H}N_{\rm H_2}^{\rm ave}w_{\rm sp}$), where $m_{\rm H}$ is the mass of a hydrogen atom, and $\mu$ is the mean molecular weight per hydrogen molecule, 2.7 \citep{Cox2000}.

Regarding the velocity properties, for each map, we extracted the average spectrum from the same region where we derived $T_{\rm peak}^{\rm ave}$ (see Section~\ref{R:COimage}). Then, we performed Gaussian fittings to determine the systemic velocity, $V_{\rm sys}$. Subsequently, we extracted spectra from each pixel along the spine and shifted the central velocity to $V_{\rm sys}$, excluding the velocity broadening due to their systematic velocity differences along the spine. We then averaged the spectra using the shifted profiles along the spine and performed another Gaussian fit to derive the velocity FWHM, $\Delta v_{\rm FWHM}$. It was used to estimate the virial line mass, $M_{\rm vir}^{\rm line}$ = 2$\sigma{_v}^2$/G \citep{Fiege_2000}, where $\sigma{_v}$ is the velocity dispersion ($\sigma{_v}$ = $\Delta v_{\rm FWHM}/2\sqrt{2{\rm ln}2}$). The derived physical quantities are listed in Table~\ref{tab:filament}. $M_{\rm line}$ is, on average, about a factor of 3 higher than $M_{\rm vir}^{\rm line}$, with a standard deviation of $\sim$2 for the ratio $M_{\rm line}$/$M_{\rm vir}^{\rm line}$. Given that virial mass estimates carry uncertainties on the order of a factor of three or more \citep[e.g.,][]{Heyer_2015}, and considering the lack of other reliable methods to determine mass and column density across the entire cloud at scales of $\sim$0.1\,pc, the mass estimates derived from both the X$_{\rm CO}$ factor and velocity dispersion should be regarded as complementary. Their mutual consistency supports the reliability of these estimates, ensuring at least an order-of-magnitude accuracy.

Although we do not delve deeply into the gravitational bounding due to its uncertainties, it is generally considered that the sample is self-gravitating as a whole. Note that we coordinated these estimated methods with the physical parameter measurements of LMC high-mass star-forming filament observations at a resolution similar to that in this study \citep{Tokuda_2023}, except for the different tracers used, allowing for comparative analysis (for detailed discussion, see Section~\ref{D:v_diff}).

\begin{figure}[htbp]
    \centering
    \includegraphics[width=1.0\columnwidth]{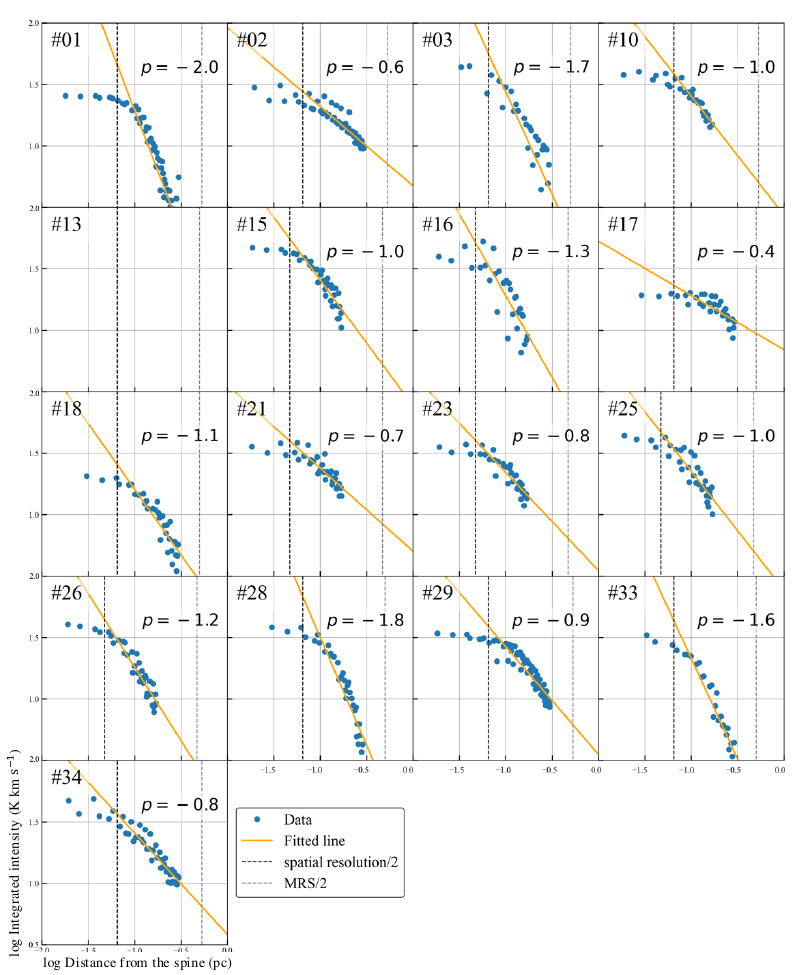}
    \caption{The average radial profile of CO intensity derived along the spines based on the FilFinder analysis. The black dashed line indicates half of the spatial resolution, and the gray dashed line indicates half of the maximum recovering scale (MRS). For data composed from multiple configurations (see Table~\ref{tab:target}), the smallest value is adopted in a conservative manner. The slope fitted to the data points between the black and gray lines is shown in orange straight lines. For \#13, no spine was identified by the FilFinder analysis.}
    \label{fig:radial_profile}
\end{figure}

We determined the connections between local peaks within the CO-emitting regions using the above procedure, but it may be premature to simply define the identified spines as filaments. To determine whether these structures are filaments or not, we conducted an additional analysis. We created average radial intensity profiles perpendicular to the spines (Figure~\ref{fig:radial_profile}). The subsequent power-low fitting derived the radial profile exponent $p$. The resulting $p$ values range from $-$0.5 to $-$2.2, with typical fitting errors of $\pm$0.1. Observational studies of interstellar filaments in low- and high-mass star-forming regions of the MW \citep[e.g.,][]{Arzounamian_2011,Palmeirim_2013,Andre_2016} and numerical works \citep[e.g.,][]{Vazquez_2019} suggest that filament profiles can be approximated by a Plummer-like function, with the radial dependence of column density (or intensity) typically following $\propto r^{-1}$. While the peak intensity of CO is likely sensitive to temperature (see Section~\ref{Dis:T_diff}), it has been noted that the column density converted from the integrated intensity using the empirical $X_{\rm CO}$ factor is not strongly dependent on the internal temperature \citep{Shetty_2011}. As a tentative criterion for this study, we classified structures with $p$ steeper than $-$1.0 as filaments and the others as non-filaments (see the last column in Table~\ref{tab:filament}). For the cases with $p$ significantly steeper than $-$1.0 (such as \#03, \#28, and \#33), there may be a contribution from their high-brightness temperatures along the spine, but at the very least, this categorization closely matches the visual impressions of the structures shown in Figures~\ref{fig:12CO32_Tpeak_view}, \ref{fig:mom0} and \ref{fig:12COexample}, and provides a reasonable threshold for statistically discussing the differences between the two categories.

\subsection{Zoom on views of CO clouds}\label{COzoom}

\begin{figure}[htbp]
    \centering
    \includegraphics[width=0.98\columnwidth]{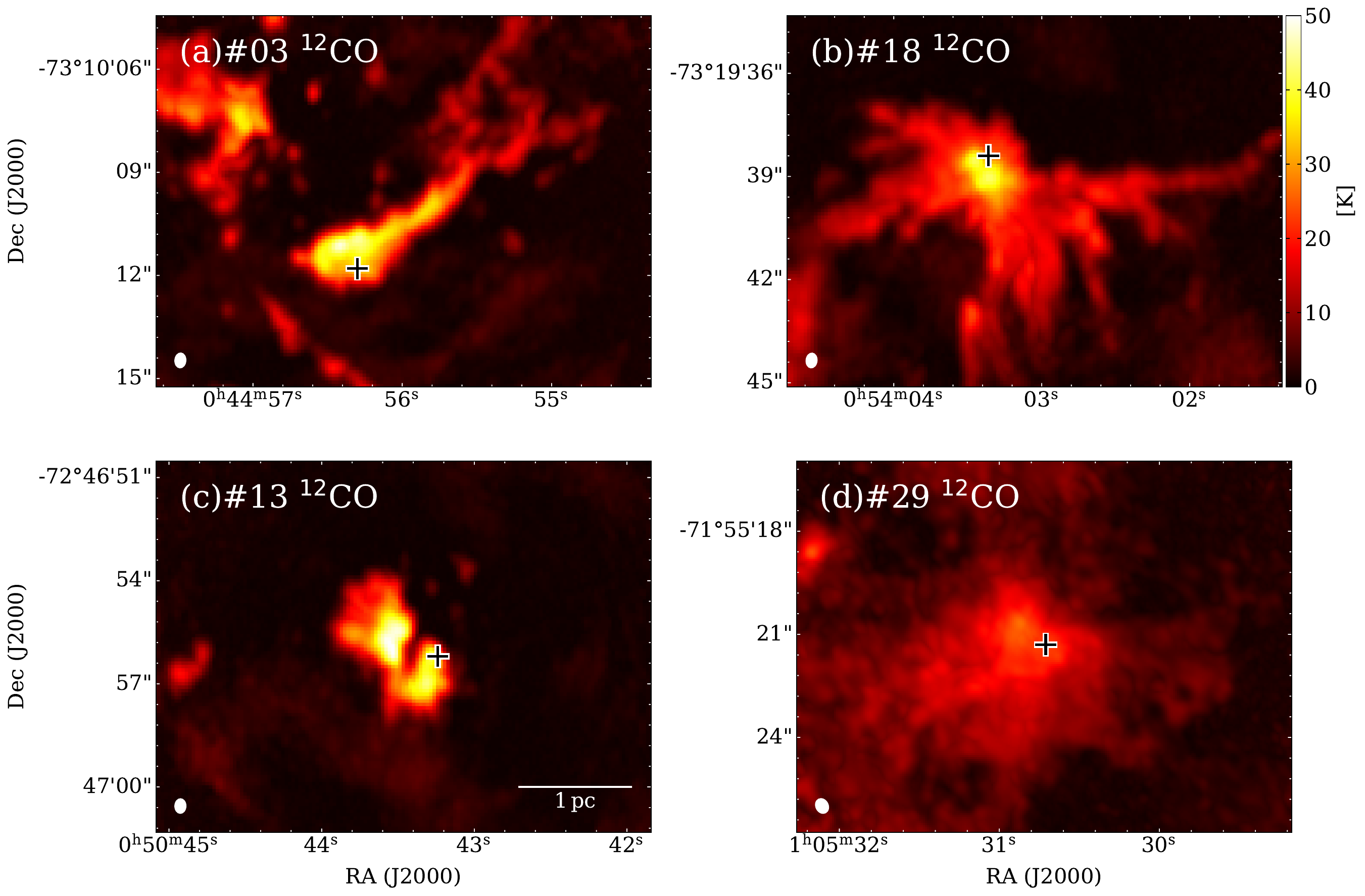}
    \caption{Enlarged views of molecular clouds associated with massive YSOs in the SMC. To illustrate the diversity of their shapes, we show (a) a single filament, (b) a hub filament, (c) a spatially compact cloud, and (d) a diffuse cloud. The color-scale images show the peak brightness temperature of CO(3--2). The white ellipses in the bottom left corners indicate the beam size. The positions of the crosses represent the massive YSOs identified by Spitzer (see Table~\ref{tab:target}).}
    \label{fig:12COexample}
\end{figure}

To more clearly illustrate the cloud distributions around protostars categorized as filaments and others, Figure~\ref{fig:12COexample} shows zoomed-in views for four fields. These were chosen because their distributions, as determined by visual inspection, are qualitatively different from each other, making them suitable for individual descriptions.

YSO~\#03, one of the targets categorized as filaments, shown in panel (a), has a structure extending approximately 3\,pc to the northwest from the YSO. The peak brightness temperature is as high as 30--50\,K, not only near the YSO position but also across the entire filamentary cloud. YSO~\#18, shown in panel (b), has an even more complex shape, with 3--4 elongated structures extending southwest of the YSO. This feature is known as a hub-filament structure, similar to those detected in the MW and the LMC \citep[e.g.,][]{Myers_2009,Tokuda_2019,Tokuda_2023,Kumar_2020}. The two sources (YSO~\#03 and YSO~\#18) have molecular outflows (see Table~\ref{tab:filament}), indicating that they are still in the active accretion phase. Other clouds with relatively simple filamentary structures, similar to the distribution in panel (a), include e.g., \#01, \#26, and \#33. CO clouds resembling the complex hub-type structure in panel (b) include e.g., \#15 and \#18. Whether single-filament or hub-filament structures are more significant in high-mass star formation remains an important topic \cite[e.g.,][]{Maity_2024,Kashiwagi_2024}. However, a strict categorisation of each filament cloud into two further types is beyond the scope of this paper, and we limit our claim to the possibility that at least a few targets may have hub-like structures.

Panels (c) and (d) show somewhat different structures. Although the peak brightness temperature of \#13 cloud is similar to the aforementioned clouds (\#03 and \#18), at 30--50\,K, it does not exhibit a high aspect ratio. It appears to be a spatially compact cloud within $\sim$1\,pc. This source is the only target for which the spine was not identified in our FilFinder analysis. YSO \#29 in panel (d), has a smoother overall distribution rather than sharp structures standing out. These are categorized as non-filamentary clouds with similar features, including \#02, \#17, \#21, \#23, \#29, and \#34. As demonstrated by these four fields, the characteristics of molecular clouds associated with massive YSOs observed at a spatial resolution of 0.1\,pc are highly diverse.

\section{Discussion} \label{sec:dis}

In this study, we focus on the spatial distribution of CO clouds associated with massive YSOs in the SMC. In particular, we have quantified whether their features are elongated, that is, whether they exhibit filamentary structures. The majority of the high-mass star-forming clouds exhibit indications of filamentary structures judged from their radial profile. However, some objects lack such features and appear to have more diffuse, spatially smoothed structures. In this section, we aim to quantify the differences not only in spatial distribution but also in physical conditions, and to explore the origins of this diversity.

\subsection{Are there any differences between filamentary and non-filamentary clouds?}\label{Dis:fil_diff}

\subsubsection{Velocity dispersion}\label{D:v_diff}

\begin{figure}[htbp]
    \centering
    \includegraphics[width=1.0\columnwidth]{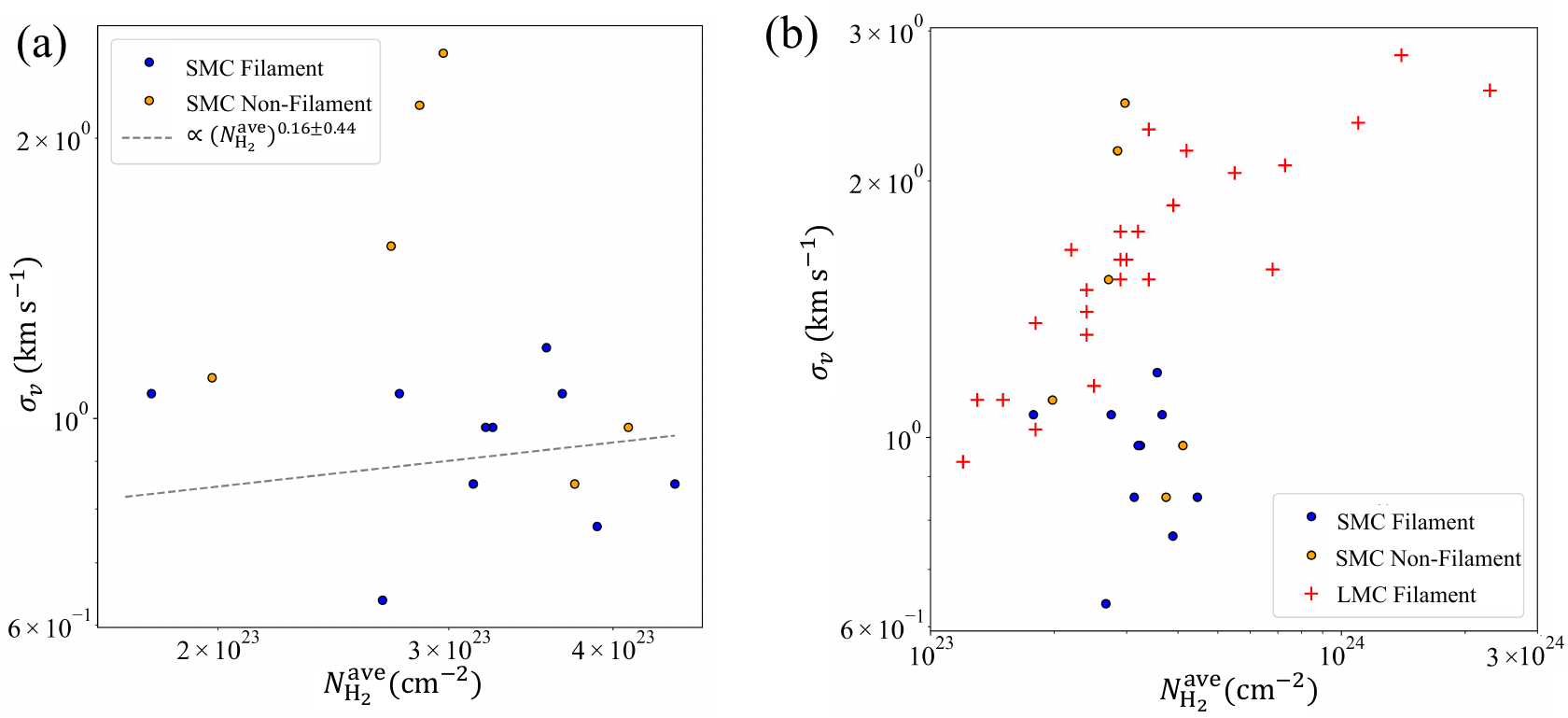}
    \caption{The relation between the average column density ($N_{\rm H_2}^{\rm ave}$) and velocity dispersion ($\sigma_{v}$) along the spine. Panel (a) visualizes the comparison between filaments and non-filaments in the SMC, while Panel (b) shows the comparison of the properties of filaments associated with massive YSOs in the LMC \citep{Tokuda_2023}. The dashed line shown in Panel (a) is the result of a least-squares fitting applied only to the filamentary clouds. Note that the leftmost data point corresponds to \#18, and since the estimate of its linewidth has a large contribution from outflow \citep{Tokuda_2022a} than the others, it is excluded from the fitting.}
    \label{fig:NH2_FWHM}
\end{figure}

Figure~\ref{fig:NH2_FWHM}(a) shows the correlation between $N_{\rm H_2}^{\rm ave}$ and $\sigma_{v}$ of the clouds associated with massive YSOs in the SMC. The most striking feature is that the objects classified as filaments tend to have a smaller velocity dispersion. This characteristic is not necessarily limited to the structures identified by the algorithm; it can also be generally observed in the moment~2 (velocity dispersion) maps shown in Figure~\ref{fig:mom2}. In studies of the MW and the LMC, it has been shown that the velocity width within star-forming filaments increases as a function of the H$_2$ column density with a power low dependence of $\sigma_v \propto N_{\rm H_2}^{0.5}$ \citep{Arzoumanian_2013,Tokuda_2023}. This relation was devised to indicate that the spatial width of the filaments remains constant at $\sim$ 0.1\,pc in various ranges of column density. The accretion of gas from outside the filament increases internal turbulence, maintaining an effective Jeans radius determined by non-thermal motion, thus allowing the filament to collapse while preserving its spatial width. Our observations do not necessarily resolve the quasi-universal value of the width of 0.1\,pc with a sufficient resolution, nor can we currently justify that the tracer is optically thin to accurately measure the column density. 
Nevertheless, the derived widths are on the order of 0.1\,pc (see Table~\ref{tab:filament}). Although the trend to increase linewidth with respect to column density is marginal in the constraints within our sample, it is not inconsistent with the above power-law relations within the range of errors (Figure~\ref{fig:NH2_FWHM}(a)).

We also focus on the differences in filament properties between SMC and LMC samples. Figure~\ref{fig:NH2_FWHM}(b) shows the $\sigma_v$--$N_{\rm H_2}^{\rm ave}$ relation both the SMC and the LMC. In the LMC, the characteristic relation, $\sigma_v \propto N_{\rm H_2}^{0.5}$ holds up to a higher column density regime (see also detailed descriptions in \citealt{Tokuda_2023}). The higher column density ($\gtrsim$10$^{24}$\,cm$^{-2}$) samples in the LMC are due to the presence of more massive and luminous YSOs. A notable feature is that the velocity dispersion of the filaments is systematically lower in the SMC filamentary clouds. In low-metallicity environments like the SMC, it has been known from global-scale CO surveys that the typical size-velocity width relation is smaller by a factor of 1.5--2 compared to those in the LMC and the MW \citep[e.g.,][]{Bolatto_2008,Ohno_2023,Saldano_2023,Saldano_2024}. This suggests that the differences in properties observed at the molecular cloud scale are also present at the star-formation-related filamentary cloud scale.

\subsubsection{Peak Brightness Temperature}\label{Dis:T_diff}
\begin{figure}[htbp]
    \centering
    \includegraphics[width=0.7\columnwidth]{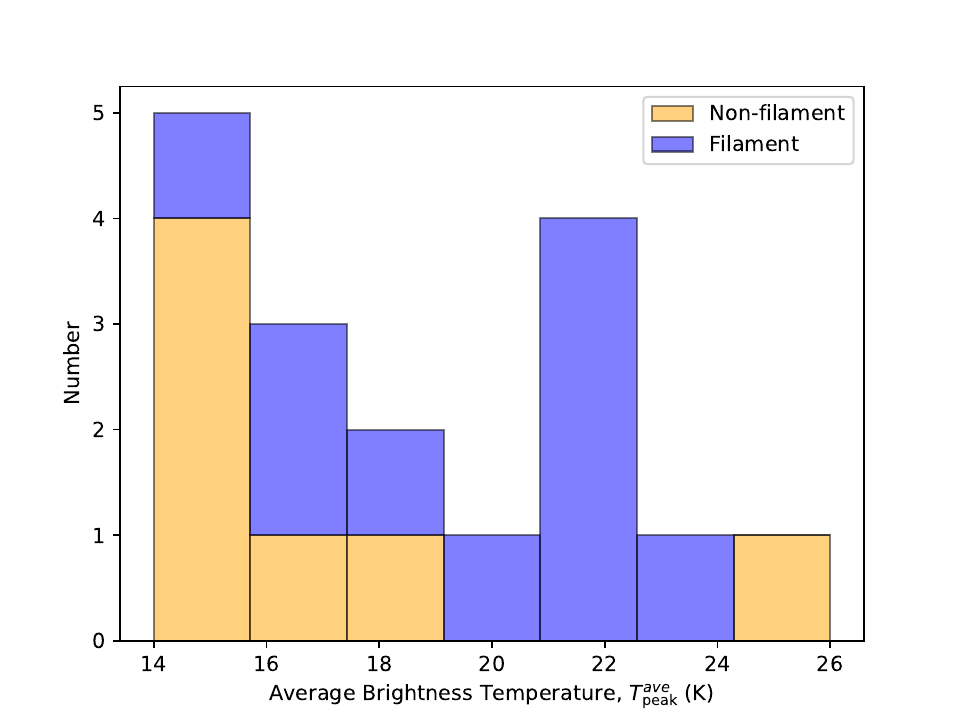}
    \caption{The histogram of the average brightness temperature of filamentary (purple) and non-filamentary (yellow) clouds.}
    \label{fig:temperature_hist}
\end{figure}

We present the differences in peak brightness temperature between the filamentary and non-filamentary clouds. The peak temperature of CO(3--2) can trace the surface temperature of the molecular cloud if it is optically thick. In the SMC dense clouds, single-dish studies obtained the results that $^{12}$CO emission is optically thick based on the line ratio of $^{13}$CO/$^{12}$CO \citep[e.g.,][]{Bolatto_2003,Nikolic_2007}. As seen in Figures~\ref{fig:12CO32_Tpeak_view} and \ref{fig:mom0}, the regions around the identified filaments (spines) have peak brightness temperatures significantly higher than 10\,K. Given that the beam filling factor is sufficiently high (close to 1), it can be assumed that the kinetic temperature of these regions is similarly high. We have divided the average temperatures listed in Table~\ref{tab:filament} into the filament and non-filament categories and plotted them as a histogram (Figure~\ref{fig:temperature_hist}). It is evident that the filament category tends to have higher brightness temperatures, while the non-filament category tends to have lower temperatures. Among the non-filaments, the one in the \#13 field has a remarkably high temperature of 26~K. As can be seen in Figure~\ref{fig:SMC_Herscchel}, most of the present targets are distributed in Herschel-bright regions (shown in cyan), where there is a large amount of cold interstellar material. However, the \#13 field is located in a bright red area, indicating that there is an optical H$\;${\sc ii} region, suggesting that it may be more affected by external heating from outside star formation than the others. In fact, the dust temperature map with a resolution of approximately 10-20\,pc \citep{Gordon_2014,Takekoshi_2017} shows a temperature of $\sim$26--32\,K around the YSO position, which is slightly higher than that in other regions.
This indicates that not only the surrounding diffuse gas but also the dense CO clouds are being heated due to an environmental effect, at least toward this source. 

In summary, the clouds with smoother distributions, which lack filamentary structures, have larger velocity widths. Comparing the filamentary structures between the Magellanic Clouds, we found that those in the SMC tend to have smaller velocity widths. Although there are some exceptions, the general trend across all observed fields is that the clouds with filaments have higher temperatures, while those without filaments have lower.

\subsection{Origin of the Diversity in the Presence or Absence of Filaments in the SMC}\label{Dis:fil_form}

\subsubsection{Did Filamentary Structures Emerge with High-Mass Star Formation?}\label{D:ealyFilament}

It has become evident that $\sim$60\% of the clouds in this study exhibit filamentary structures while others do not show such prominent features, and there are differences in the overall physical conditions, such as velocity dispersion and temperature, between the two categories. This may not be solely due to intrinsic physical differences but could also be attributed to differences in inclination angle. Even if all the clouds were filamentary, if some of their major axis are aligned along the line of sight, the filamentary features would be difficult to distinguish observationally. However, if cylindrical structures are randomly oriented on the plane of the sky, the probability of observing those oriented close to their end faces decreases \cite[c.f.,][]{Arzounamian_2011}. Furthermore, if all filaments intrinsically have similar velocity gradients, ones aligned parallel to the line of sight are expected to show steep velocity gradients, but at least from the moment-1 maps (see Figure~\ref{fig:mom1}), such trends are not observed. Therefore, we proceed the discussion under the assumption that the differences between filamentary and non-filamentary structures are physical in nature rather than observational effects.

One of the first naive questions is whether filaments emerged as initial conditions at the onset of high-mass star formation in all cases. It is necessary to rapidly form a massive dense core that eventually collapses into a high-mass star \citep{Zinnecker_Yorke_2007,Fukui_2021rapid}. A plausible formation mechanism is the interstellar magnetohydrodynamical shock compression driven by supersonically colliding flows \citep[e.g.,][]{Inoue_2013,Inoue_2018,Fukui_2021rapid}. Such colliding flows, aided by magnetic fields, can form filaments, including very massive ones whose line mass is one or two orders of magnitude higher than the thermal critical line mass, $\sim$10--20\,$M_{\odot}$\,pc$^{-1}$ at a typical cloud temperature of $\sim$10\,K \citep[e.g.,][]{Inutsuka_1992,Inutsuka_1997}. According to numerical simulations, colliding clouds with a relative velocity exceeding tens of km\,s$^{-1}$ can form filamentary structures with line masses exceeding 10$^2$\,$M_{\odot}$\,pc$^{-1}$ \citep{Abe_2021}, which is one order of magnitude greater than those typically observed in low-mass star-forming regions \cite[e.g,][]{Andre_2014}. Such extremely massive filaments appear to have multiple sub-filaments, i.e., hub-filament structures \citep[e.g.,][]{Maity_2024}. The central ridge filament with the highest column density eventually fragments into several $\sim$100\,$M_{\odot}$ cores, leading to the formation of high-mass stars \cite[e.g.,][]{Tokuda_2019,Fukui_2019,Shimajiri_2019}. The filaments we identified in this SMC study also have high-line masses of 10$^2$--10$^{3}$\,$M_{\odot}$\,pc$^{-1}$ (see Table~\ref{tab:filament}), consistent with the theoretical predictions expected from the above-mentioned colliding flows scenario. In addition to the presence of such massive filaments serving as indirect evidence of these drastic events, observational hints of colliding flows are emerging in the SMC based on H$\;${\sc i} \citep{Nigra_2008,Fukui_2020NGC602} and CO cloud observations \citep{Neelamkodan_2021}. In any case, it is reasonable to consider that once these extremely massive filaments have formed, high-mass star formation naturally follows inside them.

\subsubsection{Filament evolutionary stage based on the cloud characteristics and associated YSOs}\label{D:Filament_ange}

This subsection discusses the ages of the filaments. Determining the exact age of the filaments themselves is challenging, but we can gain indirect hints from the properties of their associated YSOs. An important observational fact is that some objects with evident filamentary-structures are associated with outflows (\#03, \#18, and \#33), suggesting that the protostars in these regions are in an earlier evolutionary stage, whose time scale is on the order of 10$^4$\,years after the protostar formation judging from the dynamical time of the flows. Only YSO~\#34 is an exception, showing its outflow but a non-filamentary shape. Unlike in more metal-rich environments, where CO can trace comparatively lower-density gas, CO in the SMC primarily probes high-density ($\gtrsim$10$^{4}$\,cm$^{-3}$) regions (see Section~\ref{sec:intro}). As a result, the dense and active outflows observed here presumably capture the earliest phases of massive star formation, characterized by the most vigorous mass accretion. In summary, the filamentary clouds tend to be accompanied by protostellar outflows, implying that they may be in a younger evolutionary stage than non-filamentary clouds.

Examining the morphological characteristics of the CO cloud can also offer insights into how much time has passed since the massive star's formation. If there were compact H\;{\sc ii} regions around a massive star with an age of $\gtrsim$10$^5$ years, we would expect to see $\sim$1\,pc diameter holes in CO \citep{Saigo_2017} and pillar structures along the rims, as observed in the LMC N159 Papillon Nebula \citep{Fukui_2019}. However, in the current sample of 17 YSOs, we do not observe such clear features. Therefore, the present targets are likely in an evolutionary phase that begins at $\sim$10$^4$ years, when the massive YSOs drive outflows, and extends up to $\sim$10$^5$ years, by which time the parental molecular cloud would be disrupted by the development of an H$\,${\sc ii} region. This timeframe provides a mild constraint on the age of the observed filament after the high-mass star formation.

\subsubsection{Did Filamentary Structures Become Less Prominent During Star Formation?}\label{D:lateFilament}

To explain why some observed clouds exhibit filamentary structures while others do not, we consider the possibility that filaments gradually lose their defining features with time after their initial formation. We envisage a three-stage time evolution: (Stage~1) filament formation leads to an increase in temperature, (Stage~2) due to inefficient cooling, the high-temperature state persists for a long time before gradually decreasing and turbulent motion develops, and (Stage~3) turbulence smoothens the filamentary structures or environmental effects originating from the SMC prohibit further filament development. The following paragraphs explain this scenario one by one.

{\bf Stage~1:} In the early phase of filament formation, it may be supported by the accretion or compression of external diffuse gas. This process helps maintain the filament's width as discussed in Section~\ref{D:v_diff}. Furthermore, the higher temperatures observed in filamentary clouds may also support the idea that filaments are younger than the others and have formed recently. The gas collision with different velocities can result in shock heating, leading to high brightness temperatures of CO(3--2), i.e., high kinematic temperatures \citep[e.g.,][]{Tokuda_2018}. Another heating mechanism in low-metallicity environments may be the formation of H$_2$ molecules, which becomes efficient at relatively high density (10$^4$--10$^5$\,cm$^{-3}$) regions \citep{Omukai_2005,Chon_2022} of molecular clouds. The metal-poor cooling inefficient condition provides high-detectable probability in high-temperature states, with average temperatures exceeding 20\,K in some of the filamentary clouds (see Table~\ref{tab:filament} and Figure~\ref{fig:temperature_hist}). If the deformation of the shock fronts that formed the filaments is the origin of the turbulence \citep{Kobayashi_2020}, then the maintenance of the high-temperature state (i.e., high thermal pressure) can counteract shock compression or delay the generation of turbulence \citep{Chon_2021,Kobayashi_2023}. This could explain why the filaments in the SMC have smaller velocity dispersions compared to those in the LMC (Figure~\ref{fig:NH2_FWHM}(b)). Although the SMC's metallicity is only about a factor of two lower than that of the LMC, there may be a very sensitive transition occurring within that range.

{\bf Stage~2:} The next phase in the time evolution is likely driven by a decrease in temperature. In the theoretical and MW observational studies, filament formation is thought to arise from shock compression driven by large-scale flows \cite[e.g.,][]{Inoue_2013,Andre_2014,Inutsuka_2015}. While such shocks may briefly raise the temperature, abundant metals enable rapid cooling that restores the filaments to their lower-temperature conditions \citep{Koyama_2000}. As a matter of fact, in Galactic infrared-dark clouds as high-mass star-forming sites with H$_2$ densities, observed scales, and resolutions similar to those in our study, the derived temperatures are on the order of $\sim$10\,K \citep{Busquet_2013}. By contrast, if a similar shock-induced filament formation with a temperature increase operates in the more metal-poor SMC, the reduced availability of coolants allows the gas to remain at higher temperatures as discussed in the previous paragraph. Once the temperature drops to around 10~K, the thermal behavior will become similar to that observed in the LMC and the MW, where the thermal pressure no longer counteracts the turbulence, leading to an increase in velocity dispersion. This is consistent with the enhanced velocity dispersions observed in the non-filamentary clouds compared to the filamentary ones as shown in Figure~\ref{fig:NH2_FWHM}. 

{\bf Stage~3:} Once the filament-forming inflow ceases, the filamentary structures can be smoothed out due to the expansion of the shock-compressed layer, as shown by numerical simulations \citep{Abe_2022}. If this process progresses on the crossing timescale of the turbulence, with a typical velocity width of 1\,km\,s$^{-1}$ and a filament width of 0.1--0.2\,pc (see Table~\ref{tab:filament}), it would take approximately 10$^5$ years. This timescale is consistent with the filamentary cloud timescale as discussed in Section~\ref{D:Filament_ange}. Alternatively, the characteristics of the flows forming filaments in the SMC differ from those in the high-mass star-forming regions of the LMC and other galaxies. For example, the duration of the filament-forming inflows might be shorter because of the shortage of gas budget in the SMC \citep[e.g.,][]{Mizuno_2001,Stanimirovic_1999}. In the LMC, high-velocity H$\;${\sc i} flows with a long path-length caused by the last galactic interaction \citep{Fukui_2017,Tsuge_2019,Tsuge_2024} contribute to globally ordered filamentary structure across various regions \citep{Fukui_2019,Tokuda_2019,Tokuda_2022a}. In spiral galaxies such as the MW and M33, very long ($\sim$100\,pc) molecular filaments are observed along the spiral arms \citep[e.g.,][]{Jackson_2010,Tokuda_2020,Kohno_2022}, suggesting that coherent flows along the stellar potential may help to maintain filamentary structures. In the metal-poor, irregular dwarf galaxy SMC, not only thermal effects but also several environmental factors should prevent filamentary structures from persisting even if they have formed.

\subsection{Implications for Interstellar Medium and Star Formation in Low-Metallicity Environments}\label{Dis:impact}

Based on the insights of the present study, we discuss the behavior of the interstellar medium and star formation in the SMC as a low-metallicity template. Previous global CO surveys have revealed some physical differences compared to metal-rich regions, such as the size-linewidth relation, where the velocity dispersions at a given cloud radius are smaller compared to those in the MW and LMC \citep{Bolatto_2008,Ohno_2023,Saldano_2023,Saldano_2024}. If molecular clouds are formed by converging flows of atomic hydrogen clouds, one explanation for the smaller velocity dispersion in low-metal environments could be reduced cooling efficiency. The turbulent energy injected by the converging flows could be counteracted by the thermal pressure of the warm gas as discussed in Section~\ref{D:lateFilament}. Numerical simulations demonstrate this behavior in the formation of the cold neutral medium just before the formation of molecular clouds \citep[e.g.,][]{Kobayashi_2023}. Investigating molecular clouds, their precursor gas, and the thermal evolution of these structures in the SMC can therefore provide valuable insights into the behavior of the interstellar medium in low-metallicity environments, where these characteristics are only just beginning to be understood.

Unlike in more metal-rich regions, the metal-poor and irregular nature of the SMC suggests that star formation activity may behave differently than in solar-metallicity environments. Although the region of the SMC examined in this study predominantly consists of dense and massive filaments as a progenitor of high-mass stars, it is likely that lower column density clouds in the outskirts and other regions can form low-mass stars. Understanding the behavior of these lower-mass-forming clouds is particularly important for gaining a comprehensive view of the overall star formation activity in such metal-poor conditions. In studies of star-forming regions in the solar neighborhood, it has been suggested that fragmentation along the axis of filaments promotes the formation of solar-mass dense cores \citep[e.g.,][]{Shimajiri_2023}, or that the formation of sub-solar-mass cores requires a further increase in density by the radial collapse of filaments \citep{Tokuda_2019MC5}. As suggested in our study of the SMC, the inefficient cooling and the difficulty of maintaining filamentary shapes may reduce the possibility of forming such small cores by filament fragmentation. This prediction is consistent with the observation of a top-heavy (or bottom-light) IMF in low-metallicity environments (see Section~\ref{sec:intro}).
Whether all filamentary clouds become less prominent before the clouds dissipate over time remains to be investigated. 

In addition to the YSOs confirmed by Spitzer spectroscopy studied here, many YSO candidates in the SMC with high reliability based on the spectral energy distribution fittings \citep[e.g.,][]{Sewilo_2013} have been found to be associated with compact CO clouds \citep{Ohno_2023}. These targets will also need to be included in future statistical studies by performing follow-up higher-resolution observations with ALMA to extend the parameter space covering the wider stellar spectrum. Furthermore, since our present study is based on single-line measurements of CO(3--2), it is needed to investigate the cloud properties using optically thinner isotope lines, such as $^{13}$CO and C$^{18}$O, and different transitions to more accurately constrain the column density and temperature distribution of the filamentary and non-filamentary cloud candidates. Future studies using the James Webb Space Telescope (JWST) to measure the detailed IMF down to the low-mass regime \citep[e.g.,][]{Olivia_2023,Hadel_2024}, combined with ALMA's ability to probe the physical properties of the parent molecular gas, will be crucial to deepening our understanding of star formation in low-metallicity environments.

\section{Summary}\label{sec:summary}

We collected the ALMA archival data for a total of 17 fields centered on YSOs, corresponding to approximately 30\% of the massive YSOs previously confirmed by Spitzer infrared spectroscopy in the SMC, and revealed the associated molecular clouds at a spatial resolution of $\sim$0.1\,pc. In the SMC, CO serves as a sufficient high-density tracer for regions with densities above 10$^4$\,cm$^{-3}$, allowing comparisons with dense filaments directly linked to high-mass formation in the MW and LMC. The key findings are summarized as follows:

\begin{itemize}
\item[1.] The spatial distribution of CO(3--2) qualitatively exhibits a diversity of structures, ranging from filaments to more extended non-filamentary structures. To characterize these visual impressions more quantitatively, we identified elongated structures from the CO distribution using the FilFinder algorithm. We classified eleven clouds with radial profiles steeper than $r^{-1}$ from the spine center as filaments, while the other six clouds were classified as non-filaments. The line masses of the filaments range from 10$^2$ to 10$^3$\,$M_{\odot}$\,pc$^{-1}$, consistent with those reported in the MW and LMC high-mass star-forming regions.

\item[2.] We found systematic differences in the physical properties of filamentary and non-filamentary clouds. The former tend to have smaller velocity dispersions relative to their column densities and exhibit higher temperatures. For clouds with filamentary morphology, the velocity width tends to increase monotonically with column density, consistent with the relationship observed in the LMC. However, the velocity widths of the filaments in the SMC are generally smaller than those in the LMC.

\item[3.] 
The high temperatures observed in the filaments suggest that they likely preserve the heated conditions related to their cloud formation. In addition, YSOs with protostellar outflows have been found in some filamentary clouds. In contrast, the non-filamentary clouds do not clearly exhibit these characteristics, suggesting they might be at a later evolutionary stage. This suggests that high-mass star-forming filaments may become less prominent at an earlier phase before the parental cloud dissipation. This behavior has not been reported in the MW and LMC. Understanding this phenomenon is crucial for the study of star formation in low-metallicity environments such as the SMC, as it highlights differences in the thermal evolution and dynamics of colliding flows that are important for determining the universality or metallicity dependence of the IMF. 

\end{itemize}

\begin{acknowledgments}
We would like to thank the anonymous referee for useful comments that improved the manuscript. This paper makes use of the following ALMA data: ADS/JAO. ALMA\#2019.1.00534.S, \#2021.1.00518.S and \#2019.1.01770.S. ALMA is a partnership of ESO (representing its member states), the NSF (USA), and NINS (Japan), together with the NRC (Canada), MOST, and ASIAA (Taiwan), and KASI (Republic of Korea), in cooperation with the Republic of Chile. The Joint ALMA Observatory is operated by the ESO, AUI/NRAO, and NAOJ. This work was supported by a NAOJ ALMA Scientific Research grant Nos. 2022-22B, Grants-in-Aid for Scientific Research (KAKENHI) of Japan Society for the Promotion of Science (JSPS; grant No. JP18H05440, JP19K14760, JP20H05645, JP21H00049, JP21H00058, JP21H01145, JP21K13962, and JP23H00129). Y.Z. acknowledges the support from the Yangyang Development Fund. The material is based upon work supported by NASA under award number 80GSFC24M0006 (M.S.). M.S. acknowledges partial support from the NASA ADAP Grant Number 80NSSC22K0168.
\end{acknowledgments}

\appendix

\renewcommand{\thefigure}{A\arabic{figure}}
\setcounter{figure}{0} 

\section{CO outflow identifications and moment maps}\label{A:COout_moms}

\begin{figure}[htb!]
    \centering
    \includegraphics[width=0.7\columnwidth]{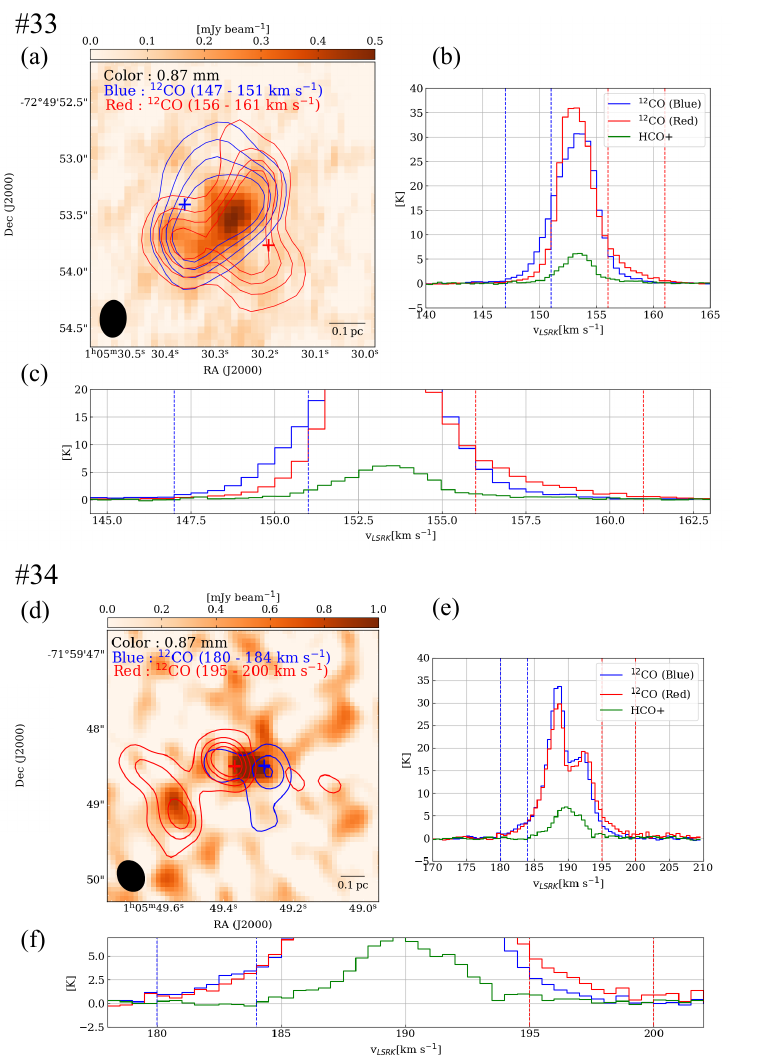}
    \caption{The 0.87\,mm continuum and CO(3--2) high-velocity emission toward Spitzer sources YSO~\#33 (a--c) and YSO~\#34 (d--f). {\bf (a)} Color-scale image shows the 0.87\,mm continuum emission toward YSO~\#33. The ellipse at the lower-left corner shows the beam size. Blue and red contours show the CO(3--2) high-velocity wing emission. The integrated velocity ranges are 147--151\,km s$^{-1}$ for the blue lobe and 156--161\,km\,s$^{-1}$ for the red lobe as indicated in the dotted lines in panel (b, and c). The lowest and subsequent contour steps are 6\,K\,km\,s$^{-1}$. 
    {\bf (b)} The blue and red profiles show the CO blueshifted and redshifted high-velocity emission, respectively, toward YSO~\#33 extracted from a 0\farcs45 radius centered at red/blue crosses in panel (a). The green profile shows the HCO$^+$(4--3) spectrum at the 0.87\,mm continuum peak. 
    {\bf(c)} Enlarged views of the outflow wing in panel (b). {\bf (d--f)} Same as those in panel (a--c), but for YSO~\#34. The integrated velocity ranges for the CO emission in panel (d) are 180--184\,km\,s$^{-1}$ for the blue lobe and 195--200\,km\,s$^{-1}$ for the red lobe.}
    \label{fig:outflow}
\end{figure}

Figure~\ref{fig:outflow} shows the protostellar outflow morphology toward YSO~\#33 and YSO~\#34. Note that to investigate the high-density cores surrounding the associated protostars, we also utilized HCO$^+$(4--3) and 0.87\,mm continuum data, which were observed simultaneously with CO(3--2) (see Section~\ref{sec:obs}.) For the 17 targets in this study, we searched for wings in the CO profiles with relative velocities exceeding $\sim$10\,km\,s$^{-1}$ with respect to the central velocity determined by the HCO$^+$ emission. The details of the identification procedure are described in \cite{Tokuda_2022b}. In addition to the previously reported detections in YSO~\#03 and YSO~\#18, we identified similar high-velocity wings in the CO line profiles for YSO~\#33 and YSO~\#34. The distance from the continuum peak to the outflow lobe is $\sim$0.1\,pc, with the maximum wing velocity of $\sim$10\,km\,s$^{-1}$, and the resultant dynamical timescale on the order of 10$^4$ years. This is roughly consistent with the previously reported timescale for YSO~\#18 \citep{Tokuda_2022b}.

\begin{figure}[htb!]
    \centering
    \includegraphics[width=1.0\columnwidth]{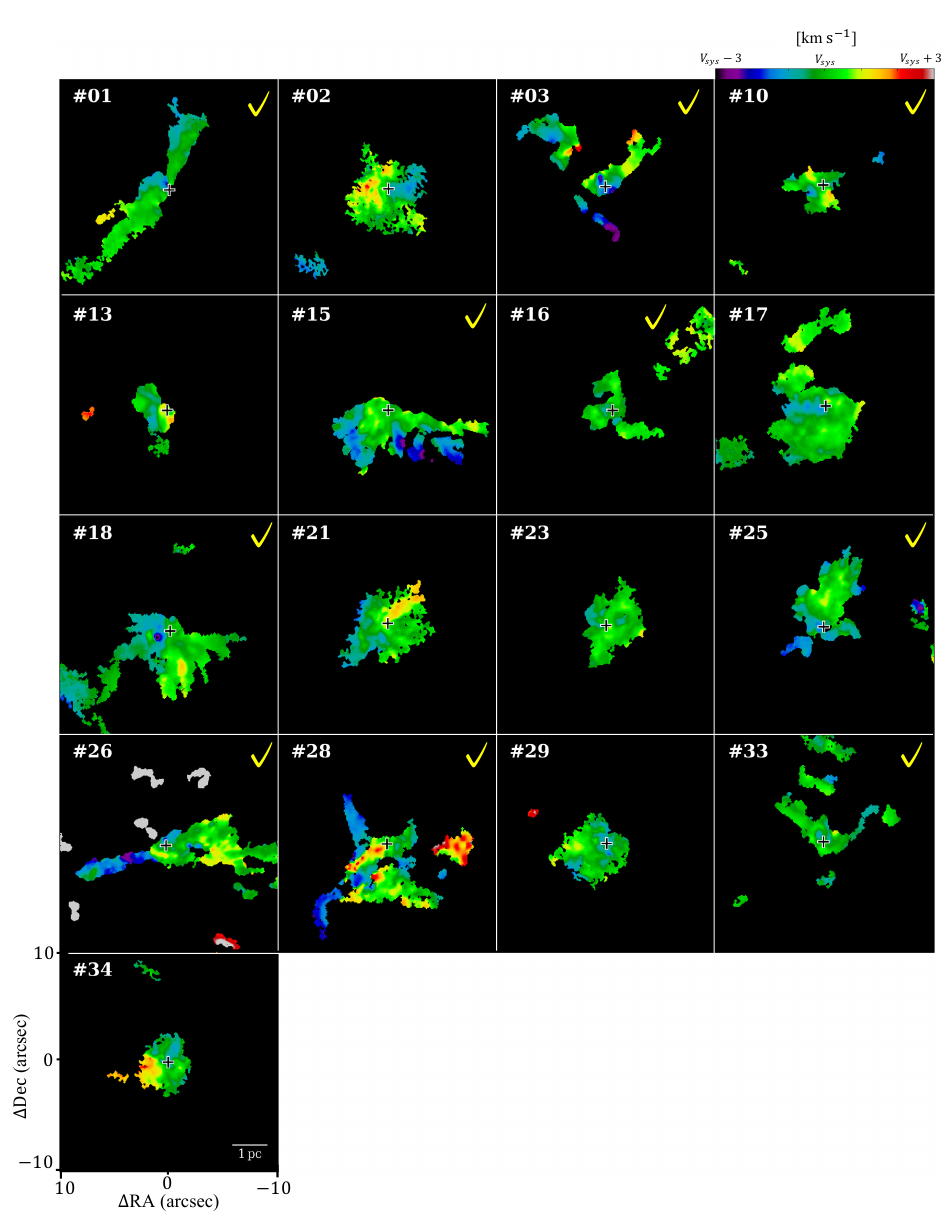}
    \caption{
    Moment~1 (the velocity field) maps of CO(3--2) toward massive YSOs from our sample.  Black plus signs show the Spitzer positions of the YSOs \citep{Oliveira_2013}, and yellow check marks indicate fields associated with filaments (see Section~\ref{R:filament_id}). The systemic velocities ($V_{\rm sys}$) are listed in Table~\ref{tab:filament}. See Figure~\ref{fig:mom0} for moment~0 maps.}
    \label{fig:mom1}
\end{figure}

\begin{figure}[htb!]
    \centering
    \includegraphics[width=1.0\columnwidth]{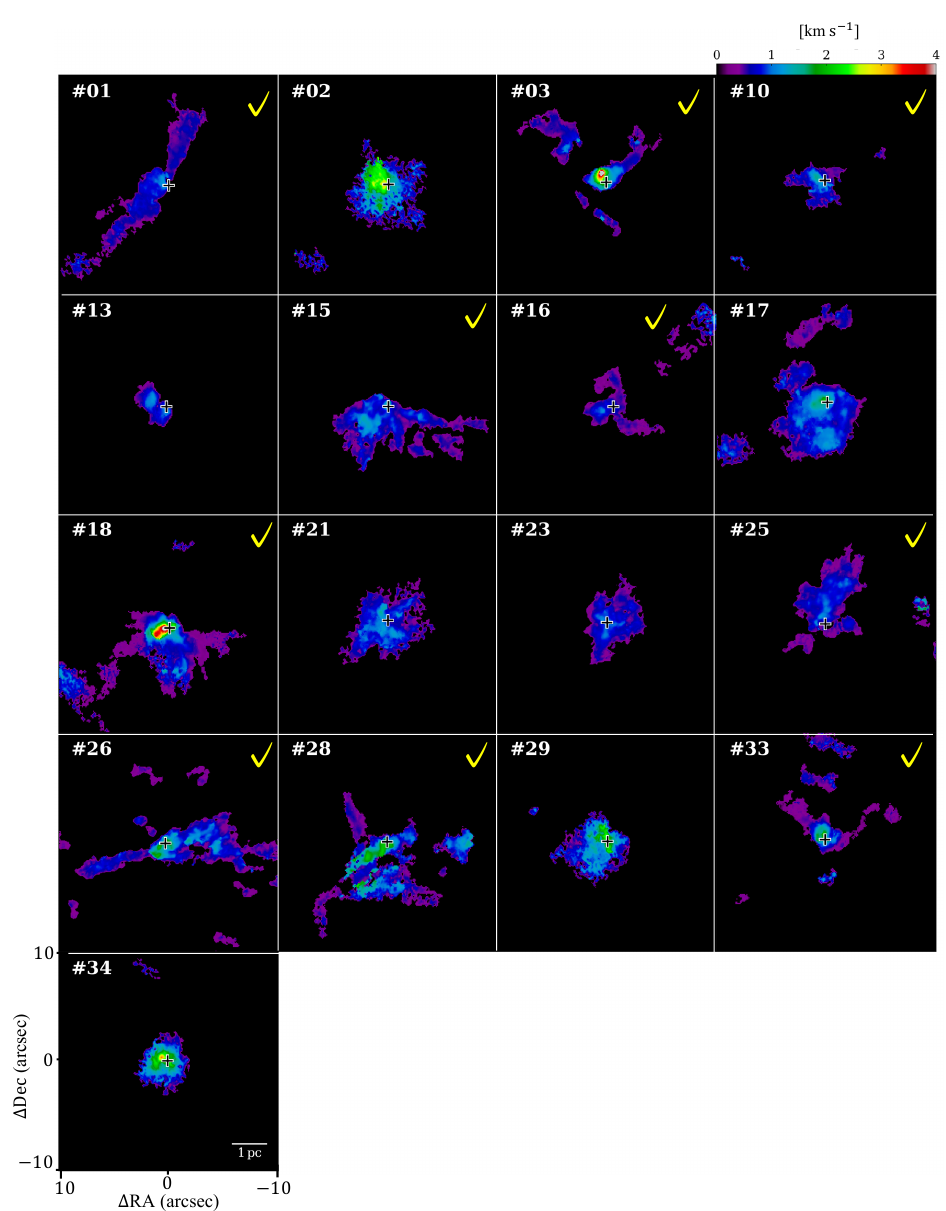}
    \caption{
    Moment~2 (the velocity dispersion) maps of CO(3--2) toward massive YSOs from our sample.  The symbols are the same as in Figure~\ref{fig:mom1}.}
    \label{fig:mom2}
\end{figure}

Figure~\ref{fig:mom1} shows the moment~1 map, which represents the velocity field of the CO clouds. Figure~\ref{fig:mom2} displays the moment~2 map, which indicates the velocity dispersion.

\software{astropy \citep{Astropy18}}

\end{document}